\documentclass[journal]{IEEEtran}

\usepackage{xcolor,soul,framed} 

\colorlet{shadecolor}{yellow}
\usepackage[pdftex]{graphicx}
\graphicspath{{../pdf/}{../jpeg/}}
\DeclareGraphicsExtensions{.pdf,.jpeg,.png}

\usepackage[cmex10]{amsmath}
\usepackage{array}
\usepackage{mdwmath}
\usepackage{mdwtab}
\usepackage{eqparbox}
\usepackage{url}
\usepackage{amsmath}
\usepackage{amssymb}
\usepackage{booktabs} 
\usepackage{caption}  
\usepackage{array}    
\usepackage{tabularx}  
\usepackage{textcomp}  

\newcolumntype{L}{>{\raggedright\arraybackslash}X}
\usepackage[american]{circuitikz}
\usetikzlibrary{positioning} 

\ctikzset{bipoles/thickness=1}
\tikzset{
  block/.style={
     draw, thick, minimum width=1.6cm, minimum height=0.9cm,
     font=\small\sffamily, align=center
  },
  lbl/.style={font=\footnotesize\sffamily}
}

\begin{document}
\bstctlcite{IEEEexample:BSTcontrol}
    \title{ Integrated Sliding-Short/Probe Tuner with Doorknob Transition for High Quality-Factor Cavities }
  \author{Saptarshi Biswas and 
      Sven G. Bilén,~\IEEEmembership{Senior Member,~IEEE}\\

  \thanks{S. Biswas, Post-Doctorate Researcher, Oregon State University, Corvallis, OR, 97331 USA (email:saptarshi.biswas@oregonstate.edu).}
  \thanks{Sven G. Bilén, Professor, The Pennsylvania State University, University Park, PA 16802, USA (email:sbilen@psu.edu).}%
}


\maketitle

\begin{abstract}
We present a compact impedance tuner integrated into the launch adapter of a waveguide-fed, high quality-factor cavity. The launcher combines a sliding short, a doorknob transition, and a micrometer-adjustable coaxial probe so that impedance matching and external coupling are controlled at the cavity interface. A transmission-line chain-matrix model maps the mechanical settings to the input match and the loaded quality factor, including the feedthrough capacitance. Full-wave simulations and measurements on a Ku-band prototype validate the synthesis, achieving approximately $30$-dB return loss near $17.7$--$18.1$~GHz and about $0.8$-dB insertion loss in the upstream feed network. Simulations indicate peak nozzle electric fields of about $1.8\times10^5$~V/m for $1$~W incident power. A parametric study shows that long backshort offsets excite a parasitic stub resonance that produces a second return-loss minimum and localizes energy behind the transition; keeping the offset below about $0.4$ guided wavelengths avoids this. During plasma operation, \textit{in-situ} retuning compensates load drift and improves absorbed-power fraction across helium, ammonia, and hydrogen operating cases.
\end{abstract}


%
\IEEEpeerreviewmaketitle

\section{Introduction}
\IEEEPARstart{E}{fficient} power transfer at the interface between a waveguide feed and a high-$Q$ cavity is a recurring challenge in microwave and millimeter-wave hardware. In waveguide-coupled resonators, small changes in launcher geometry and cavity loading can produce large changes in the input reflection coefficient and the external coupling factor, increasing standing-wave voltages and limiting power handling. Practical systems therefore require compact tuning mechanisms that provide (i) deep matching at the input reference plane and (ii) controllable coupling and loaded quality factor for a targeted cavity resonance. These requirements arise in dielectric heating and plasma processing systems~\cite{jungATE2021,chenAS2021}, in propulsion test cavities~\cite{clemensPhD2008,micciEUCASS2009}, and in instrumentation resonators such as electron paramagnetic resonance cavities~\cite{juddAMR2021}. In all cases, mismatch wastes delivered power and can stress microwave sources~\cite{bilikCOMITE2008}, motivating impedance tuners that can be integrated directly at the waveguide--cavity interface.

Waveguide-fed cavities are traditionally matched using reactive elements inserted along the feed. Three-stub tuners are a standard solution in high-power systems~\cite{kurzrokTMTT1963,laiJMP2020}, but they are bulky and prone to breakdown~\cite{bilikCOMITE2008}. Coaxial probe feeds are also common~\cite{Asmussen,Plaza2007}, but fixed geometries operate at only one condition, while adjustable mechanisms are often mechanically delicate. Similarly, aperture coupling through irises~\cite{bethePR1944} is essentially static, requiring hardware swaps to change coupling levels.

Contemporary tuner research has refined these building blocks into synthesized microwave networks with explicit performance metrics. In the waveguide domain, precision matching has been demonstrated using sliding shorts for doorknob couplers~\cite{Franco2001} and dielectric backshorts for W-band~\cite{Kiuru2007,Stanec2013}. Advanced calibration techniques have further improved the characterization of coaxial-to-waveguide transitions~\cite{LozanoTMTT2010,LozanoTMTT2021}. Parallel efforts in planar technology have produced agile impedance tuners using MEMS switches and capacitors~\cite{Kim2001,Papapolymerou2003}, while resonant impedance tuners based on coupled-resonator synthesis now offer wide Smith-chart coverage with defined power handling~\cite{Shaffer2024,Roessler2023,MoranGuizan2024}.

However, most such tuners are implemented as stand-alone networks inserted between the source and load, rather than being integrated into the cavity interface itself. This integration is particularly critical for time-varying high-$Q$ loads, such as plasma-loaded resonators~\cite{Ramesh2024,Kabir2025}, pulse compressors~\cite{Jiang2021}, and anapole structures~\cite{Akram2025}, where the effective loading evolves during operation. Unlike general-purpose tuners that prioritize broad Smith-chart coverage, the objective here is cavity-centric: to realize deep matching ($\Gamma \rightarrow 0$) while directly controlling the external coupling coefficient $\beta$ and loaded quality factor $Q_{\rm L}$ using a compact waveguide--doorknob--probe interface.

This work presents a launcher-integrated tuner that eliminates external stub boxes while remaining compatible with high-power, vacuum-based operation. The contributions are as follows:
1) A compact impedance tuner that embeds a sliding short and an adjustable coaxial probe inside a doorknob transition.
2) A closed-form chain-matrix model that links mechanical settings to input reflection, $\beta$, and $Q_{\rm L}$, explicitly including feedthrough capacitance.
3) Validation via full-wave simulation and measurement, including the identification of a backshort-length constraint ($l_{\rm s} \le 0.4\lambda_{\rm g}$) that suppresses parasitic stub resonances.
4) Demonstration of \emph{in-situ} retuning during time-varying plasma operation using a 17.8-GHz microwave electrothermal thruster (MET) cavity~\cite{biswasPhD2023}. Sections~\ref{analytical model}--\ref{conclusion} present the analytical model, experimental setup, and results.
\section{Analytical Model of the Integrated Coax-Waveguide Tuner}
\label{analytical model}

\subsection{Reference planes and transmission-line scaffold}\label{sec:planes}
\begin{figure} [ht!]
  \begin{center}
  \includegraphics[width=3.7in]{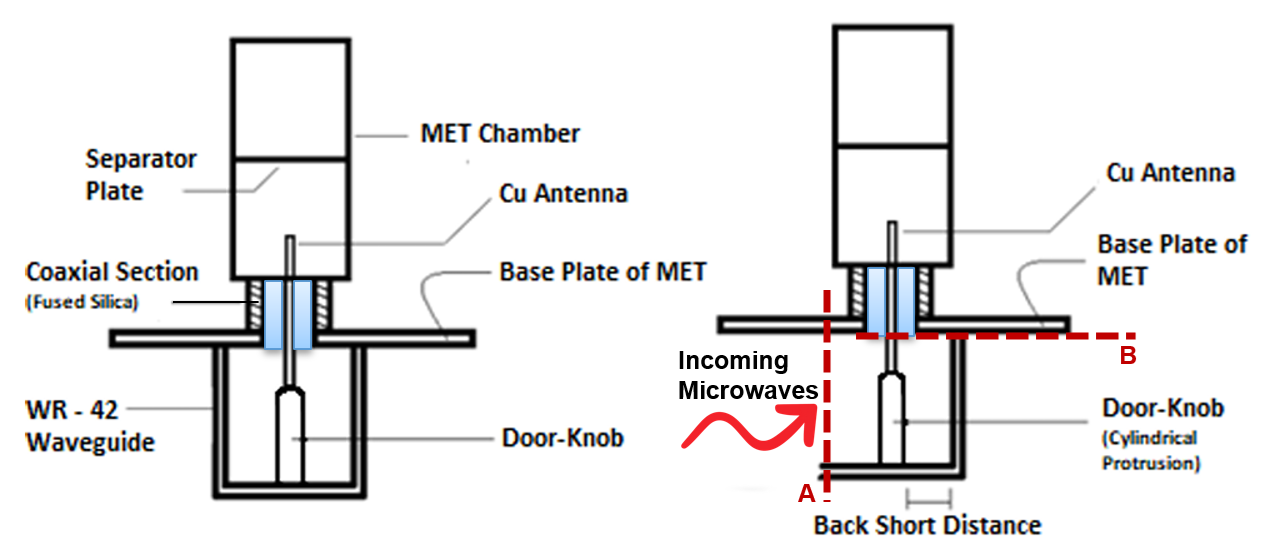}\\
  \caption{The front view (left) and the side view (right) geometry of a doorknob protrusion with a variable antenna from left to right.}\label{fig:tuner_circuit}
  \end{center}
\end{figure}

Plane A is located at the WR--42 flange. The port reference impedance is the dominant-mode (TE$_{10}$) wave impedance of the air-filled guide, $Z_{0,\mathrm{wg}}(f)$, evaluated at the frequency of interest. Plane B lies at the interior cavity wall where the copper antenna enters the MET chamber 
(Fig.~\ref{fig:tuner_circuit}). 

This section develops a chain-matrix model for the integrated tuner. Similar to the resonant impedance tuners in \cite{Shaffer2024,Roessler2023}, the goal is to map a small set of mechanical knobs to the reflection coefficient and external loading of a high‑Q resonator, but here the ‘network’ is the waveguide-doorknob-probe-cavity interface itself rather than a stand-alone 50-$\Omega$ tuner.

The hardware between the planes is
partitioned into four tunable two-port blocks: (i) a
doorknob impedance transformer ($L_{\rm d}$, $C_{\rm d}$, gap~$g:l_{\rm d}$),
(ii) a sliding-short waveguide stub (offset~$l_{\rm s}$), and
(iii) a fused-silica feed-through of shunt capacitance $C_{\rm fs}$, and
(iv) a movable probe of insertion depth $h$. Their ABCD
(``chain'') matrices cascade to
\begin{equation}
\label{eq:ABCDsum}
\setlength{\arraycolsep}{2.5pt} 
\begin{bmatrix} A_{\Sigma} & B_{\Sigma} \\ C_{\Sigma} & D_{\Sigma} \end{bmatrix}
=
\begin{bmatrix} A_{\mathrm{d}} & B_{\mathrm{d}} \\ C_{\mathrm{d}} & D_{\mathrm{d}} \end{bmatrix}
\begin{bmatrix} A_{\mathrm{s}} & B_{\mathrm{s}} \\ C_{\mathrm{s}} & D_{\mathrm{s}} \end{bmatrix}
\begin{bmatrix} 1 & 0 \\ j\omega C_{\mathrm{fs}} & 1 \end{bmatrix}
\underbrace{\begin{bmatrix} 1 & jX_{\mathrm{p}}(h) \\ 0 & 1 \end{bmatrix}}_{\mathbf{M}_{\mathrm{p}}(h)}.
\end{equation}
where $(l_{\rm d},l_{\rm s},h)$ are the primary geometric parameters of the integrated launcher and $C_{\rm fs}=0.06$~pF is the fused-silica sleeve capacitance at 17.8~GHz (Ku~band). In the assembled hardware, $l_{\rm d}$ is set by fabrication (and may be adjusted only by shimming), while the two \emph{in-situ} tuning knobs are $(l_{\rm s},h)$. This chain-matrix description follows the standard tuner‑synthesis approach used for multi-resonator impedance tuners and matching networks \cite{Shaffer2024,MoranGuizan2024}, but with the cavity resonator explicitly included in the load.

\subsection{Cavity load at Plane B}\label{sec:ZL}
The dominant $\mathrm{TM}^z_{011}$ discharge behaves electrically as a parallel \(R_{\rm c}\!-\!L_{\rm c}\!-\!C_{\rm c}\) resonator.  Viewed from Plane B, i.e., after the doorknob and stub have already been traversed, three elements contribute to the local impedance
\begin{equation}
\label{eq:ZL}
Z_{\rm L}(f)=
\underbrace{\frac{R_c}{n^{2}(h)}
\frac{1}{1+j\,Q_0\left(\dfrac{f}{f_0}-\dfrac{f_0}{f}\right)}}_{\text{cavity $\mathrm{TM}^z_{011}$ impedance referred through }n(h)}.
\end{equation}
The load $Z_{\rm L}(f)$ in \eqref{eq:ZL} is the Lorentzian impedance of the $\mathrm{TM}^z_{011}$ cavity referred through the transformer ratio $n(h)$. The probe self-reactance is captured separately by the series two-port $\mathbf M_{\rm p}(h)$ in the ABCD cascade of \eqref{eq:ABCDsum}, and the fused-silica
feedthrough is modeled as a shunt capacitance $C_{\rm fs}$ via \eqref{eq:ABCDsum}. Inserting \eqref{eq:ZL} into the ABCD cascade of Fig.~\ref{fig:tuner_circuit} produces the input impedance at Plane A, from which the matching conditions of Section \ref{sec:probe_dknob} are derived.

\subsection{Input impedance and reflection coefficient}
The impedance seen at Plane A and the reflection coefficient are
\begin{equation}
\label{eq:ZinGamma}
Z_{\text{in}}(f)=\frac{A_\Sigma Z_L(f)+B_\Sigma}{C_\Sigma Z_L(f)+D_\Sigma},
\qquad
\Gamma(f)=\frac{Z_{\text{in}}(f)-Z_{0,\mathrm{wg}}(f)}{Z_{\text{in}}(f)+Z_{0,\mathrm{wg}}(f)}.
\end{equation}
\subsection{Door--knob transformer, probe impedance, and sliding short}\label{sec:probe_dknob}
As Fig.~\ref{fig:tuner_circuit} shows, the copper center-conductor first traverses the doorknob gap by a length $h_{\rm g}$ and then protrudes a distance $h$ into the cylindrical  $\mathrm{TM}^z_{011}$ cavity\footnote{The short section below Plane~B lies inside the WR-42 guide and is contained inside the doorknob matrix; only the exposed length $h$ radiates into the cavity.}. The serial network on the cavity side of Plane~A therefore contains two elements in series:

\paragraph*{(i) Doorknob $L$--$C$ step with gap inductance.}
Treating the doorknob gap as an $L$--$C$ step follows standard waveguide discontinuity modeling \cite{slater_microwave_1942} and is analogous to the stepped impedance transformers used in modern tuner structures \cite{Kiuru2007,Stanec2013}.
The inductance is augmented by the center-conductor section in the gap,
\begin{equation}
\label{eq:Ld_star}
\begin{split}
L_{\rm d}^\star(l_{\rm d},h_{\rm g}) &= L_{\rm d}(l_{\rm d})+L_{\rm g}(h_{\rm g}) \\
&= \mu_0 r_{\rm d}\left[1+\frac{1}{2}\ln\left(\frac{16g}{r_{\rm d}}\right)\right]
\;+\;
\frac{\mu_0 h_{\rm g}}{2\pi}\ln\left(\frac{g}{d}\right),
\end{split}
\end{equation}
where $L_{\rm g}(h_{\rm g})=\frac{\mu_0 h_{\rm g}}{2\pi}\ln\left(\frac{g}{d}\right)$
and the capacitance remains $C_{\rm d} \simeq \pi\varepsilon_0 r_{\rm d}^{2}/g$. Here $r_{\rm d}$ is the doorknob-post radius. The transformer ABCD matrix keeps the form
\begin{equation}
\mathbf M_d=
\begin{bmatrix}1 & j\omega L_{\rm d}^\star \\[4pt] 0 & 1\end{bmatrix}
\begin{bmatrix}1 & 0 \\[4pt] j\omega C_{\rm d} & 1\end{bmatrix}\!,  
\end{equation}
so that the doorknob contributes a series reactance
$X_{\rm d}(l_{\rm d},h_{\rm g})=\omega L_{\rm d}^\star(l_{\rm d},h_{\rm g})$ and a shunt susceptance $B_{\rm d}(\omega)=\omega C_{\rm d}$ at the coupling plane. Accordingly, $C_{\rm d}$ enters the shunt-susceptance balance in the matching condition \eqref{eq:match_conditions}, not the series-reactance sum.

\paragraph*{(ii) Probe tip inside the cavity.}
Two distinct segments of the centre-conductor must be recognized:

\begin{itemize}
\item \textit{Gap segment, $h_{\rm g}$ (coaxial).}  
      This $50-\Omega$ coax section lies \emph{below} Plane B, inside
      the door‑knob gap~$g$.  Its current returns on the cylindrical
      post, so its inductance is fully captured in
      $L_{\rm d}^{\star}$; it does \emph{not} radiate into the cavity.

\item \textit{Exposed segment, $h$ (radiating).}  
      The length that protrudes \emph{above} Plane B enters the
      cylindrical TM$^z_{011}$ cavity. Near the axis, the cavity cross-section is locally quasi-planar; therefore, Slater’s on-axis short-dipole expressions provide a convenient first-order approximation for the probe self-impedance as a function of $h$ \cite{slater_microwave_1942}. Because the probe radiates into a cylindrical cavity (TM$^{z}_{011}$) rather than a uniform rectangular guide, \eqref{Xp} should be interpreted as an \emph{effective} reactance used for synthesis. Any residual reactive error is compensated by the adjustable backshort $l_{\rm s}$, and the final operating points are validated against full-wave FEM and measurements in Sec.~\ref{results}.
\end{itemize}
\begin{align}
\label{R_rad}
R_{\text{rad}}(h) &=
 \frac{30\pi^{2}h^{2}}{\lambda^{2}}
 \cdot \frac{\sin(\pi a/\lambda)}{\sqrt{1-(\lambda/2a)^{2}}}, \\[3pt]
X_{\rm p}(h) &\approx 60\left[\ln\left(\frac{2h}{d}\right) - 1\right]
 -30\pi^{2}\dfrac{h^{2}}{\lambda^2}
 \cdot \frac{\sin(\pi a/\lambda)}{\sqrt{1-(\lambda/2a)^{2}}}.
 \label{Xp}
\end{align}
(The closed-form \eqref{R_rad}--\eqref{Xp} hold for $h\!\ll\!\lambda$ and $\lambda<2a$,
i.e., operation above TE$_{10}$ cutoff.) Here $\lambda\equiv\lambda_0=c/f$ is the free-space wavelength.
These expressions are consistent with coaxial-probe measurements in circular and rectangular waveguides reported in \cite{LiangIMS1992,MacPhieTMTT1990}. Here $d$ is the probe diameter and $a$ is the WR--42 broad-wall dimension that sets the TE$_{10}$ cutoff. Matching real parts fixes the transformer ratio so that the probe extracts exactly the intrinsic cavity loss under critical coupling:
\begin{equation}
    n^{2}(h)=\frac{R_{\rm c}}{R_{\text{rad}}(h)}.
    \label{critical coupling}
\end{equation}
These $R_{\text{rad}}(h)$ and $X_{\rm p}(h)$ fully specify the probe model: $R_{\text{rad}}(h)$ fixes the coupling ratio $n(h)$ via \eqref{critical coupling}, while $X_{\rm p}(h)$ enters as a series reactance through $\mathbf M_{\rm p}(h)$ in \eqref{eq:ABCDsum}.

\paragraph*{Composite series reactance}
Because the doorknob series inductance and the radiating probe lie in series, their reactances add:
\begin{equation}
X_{\mathrm{series}}(l_{\rm d},h_{\rm g},h)=X_{\rm d}(l_{\rm d},h_{\rm g})+X_{\rm p}(h)=\omega L_{\rm d}^\star(l_{\rm d},h_{\rm g})+X_{\rm p}(h).
\end{equation}
The fused-silica feedthrough is modeled as a shunt capacitance $C_{\text{fs}}$ via $\mathbf{M}_{\text{fs}}$ in \eqref{eq:ABCDsum}; it is not included in $Z_{\rm L}$.

\paragraph*{Four-block ABCD scaffold}
With the refined inductance, the cascade at Plane A remains
\begin{equation}
    \mathbf M_\Sigma=
    \mathbf M_{\rm d}(l_{\rm d},h_{\rm g})\;
    \mathbf M_{\rm s}(l_{\rm s})\;
    \mathbf M_{\rm {fs}}(C_{\rm {fs}})\;
    \mathbf M_{\rm p}(h).
\end{equation}

\paragraph*{Sliding Backshort (Shunt Element)} The movable short closes the WR-42 guide at an offset $l_{\mathrm{s}}$ behind the doorknob/probe coupling plane. The shorted section presents the standard short-circuited input impedance
\begin{equation}
\begin{split}
    Z_{\mathrm{s}}(l_{\mathrm{s}},f) &= j Z_{0,\mathrm{wg}}(f) \tan\left(\beta_{\mathrm{g}}(f) l_{\mathrm{s}}\right), \\
    Y_{\mathrm{s}}(l_{\mathrm{s}},f) &= \frac{1}{Z_{\mathrm{s}}}
    = -j Y_{0,\mathrm{wg}}(f) \cot\left(\beta_{\mathrm{g}}(f) l_{\mathrm{s}}\right),
\end{split}
    \label{eq:Ys_backshort}
\end{equation}
where $\beta_{\mathrm{g}}(f)=2\pi/\lambda_{\mathrm{g}}(f)$ and $Y_{0,\mathrm{wg}} = 1/Z_{0,\mathrm{wg}}$. Because the backshort is connected in \emph{parallel} with the doorknob--probe branch at the coupling plane, it is modeled in the ABCD scaffold as a shunt element,
$
\mathbf{M}_{\mathrm{s}}(l_{\mathrm{s}}) = \begin{bmatrix} 1 & 0 \\ Y_{\mathrm{s}}(l_{\mathrm{s}},f) & 1 \end{bmatrix},
$
so that the quarter-wave condition $l_{\mathrm{s}}=\lambda_{\mathrm{g}}/4$ yields $Y_{\mathrm{s}} \to 0$ (an ``open'' in shunt at the coupling plane).

\paragraph*{Matching Equations} 
At $f=f_{0}$, \eqref{eq:ZL} and \eqref{critical coupling} give $Z_{\rm L}(f_{0})=R_{\mathrm{rad}}(h)$ and the probe branch $Z_{\rm p}=R_{\mathrm{rad}}(h)+jX_{\rm p}(h)$. The shunt elements at the coupling plane therefore sum to $Y_{\Sigma}=j\omega_{0}(C_{\rm d}+C_{\mathrm{fs}})-\frac{j}{Z_{0,\mathrm{wg}}}\cot(\beta_{\rm g}l_{\rm s})+\frac{1}{R_{\mathrm{rad}}(h)+jX_{\rm p}(h)}$, and with $X_{\rm d}=\omega_{0}L_{\rm d}^{\star}$ the condition
$Z_{\mathrm{in}}(f_{0})=jX_{\rm d}+1/Y_{\Sigma}=Z_{0,\mathrm{wg}}$ implies $Y_{\Sigma}=1/(Z_{0,\mathrm{wg}}-jX_{\rm d})$,
which yields \eqref{eq:match_conditions} by separating real and imaginary parts. 
\begin{equation}
\begin{gathered}
\frac{R_{\mathrm{rad}}(h)}{R_{\mathrm{rad}}^{2}(h)+X_{\mathrm{p}}^{2}(h)}
=
\frac{Z_{0,\mathrm{wg}}}{Z_{0,\mathrm{wg}}^{2}+X_{\mathrm{d}}^{2}(l_{\mathrm{d}},h_{\mathrm{g}})},
\\[10pt]
\begin{aligned}
& \omega_{0}\!\left(C_{\mathrm{d}}+C_{\mathrm{fs}}\right)
-\frac{\cot\!\left(\beta_{\mathrm{g}} l_{\mathrm{s}}\right)}{Z_{0,\mathrm{wg}}}
\\[4pt]
& \qquad
-\frac{X_{\mathrm{d}}(l_{\mathrm{d}},h_{\mathrm{g}})}{Z_{0,\mathrm{wg}}^{2}+X_{\mathrm{d}}^{2}(l_{\mathrm{d}},h_{\mathrm{g}})}
-\frac{X_{\mathrm{p}}(h)}{R_{\mathrm{rad}}^{2}(h)+X_{\mathrm{p}}^{2}(h)}
=0, 
\end{aligned}
\end{gathered}
\label{eq:match_conditions}
\end{equation}
where all quantities are evaluated at $f_{0}$ (i.e., $\omega_{0}=2\pi f_{0}$), and $X_{\rm d}(l_{\rm d},h_{\rm g})\triangleq \omega_{0}L_{\rm  d}^{\star}(l_{\rm d},h_{\rm g})$.
The first line is the conductance (real-part) matching condition at the coupling node, and the second line enforces net susceptance cancellation at $f_{0}$, summing the shunt capacitances $(C_{\rm d}+C_{\mathrm{fs}})$, the backshort susceptance, and the susceptive parts of the series doorknob/probe branch.

Equations \eqref{eq:match_conditions} play the same role as the tuning equations in resonant impedance tuners \cite{Shaffer2024}: for any given cavity load, there is a continuum of mechanical settings that satisfy the matching conditions. Because three primary variables $(l_{\mathrm{d}}, l_{\mathrm{s}}, h)$ satisfy two constraints (with $h_{\mathrm{g}}$ typically determined by setting $l_{\mathrm{d}}$), the theoretical solution set is generally a one-parameter family. In the physical realization, $l_{\mathrm{d}}$ is fixed by fabrication, collapsing the solution to a specific operating point $(l_{\mathrm{s}}, h)$ that we select by imposing the desired coupling (e.g., critical coupling, $\beta=1$).

\paragraph*{Loaded $Q$ and reflection at resonance}

At resonance, the coupling coefficient is defined in the usual way as $\beta \triangleq Q_0/Q_{\rm e}$, where $Q_{\rm e}$ is set by the coupler. The corresponding
one-port relations at resonance are 

\begin{equation}
  Q_{\rm{L}}=\frac{Q_0}{1+\beta},\qquad
  |S_{11}(f_0)|=\left|\frac{\beta-1}{\beta+1}\right|.
  \label{eq:QL_S11}
\end{equation}

\subsection{Higher-Order-Mode (HOM) Check}
To validate the single-mode assumption used in the analytical model, a waveguide-port mode decomposition in COMSOL Multiphysics was examined to quantify higher-order content in the WR-42 section behind the doorknob. Across 17.6--18.2~GHz, and even for extended backshort offsets $l_{\mathrm{s}} \ge 9$~mm, the cumulative amplitude of the first higher-order modes (TE$_{20}$, TE$_{01}$) remained at least 40~dB below the dominant TE$_{10}$ mode. This suppression is further confirmed by the close agreement between the synthesis and full-wave results: the analytic optimum derived from \eqref{eq:match_conditions} (which assumes purely single-mode propagation) predicts $|S_{11}(f_0)| < -25$~dB and agrees with full-wave FEM sweeps within $\sim 1$~dB. This quantitative consistency indicates that the combined $R_{\mathrm{rad}}(h)$ and effective $X_{\mathrm{p}}(h)$ trends are sufficient to predict the tuning behavior without explicit inclusion of modal cross-talk.


\section{Methods}\label{sec:exp}

\begin{figure}[ht!]
  \centering
  \includegraphics[width=0.48\textwidth]{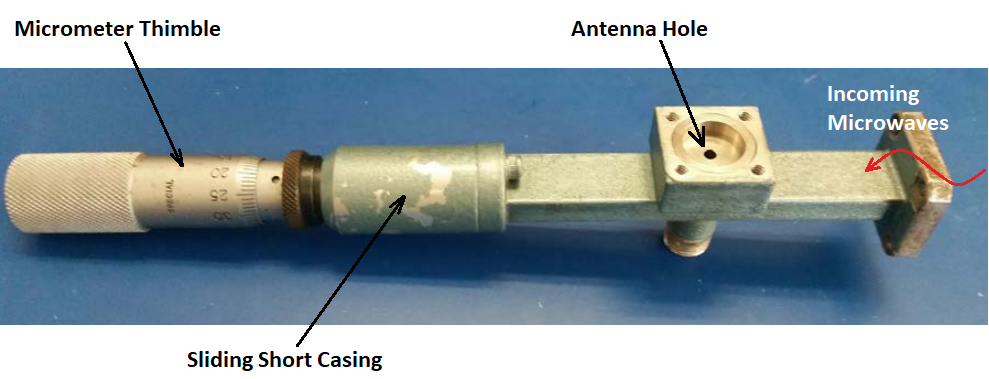}
  \caption{Waveguide cavity mount with sliding short (backshort). Incoming microwaves enter from the right (red arrow) and are reflected by the short; translating the plunger tunes the
  shunt backshort admittance $Y_{\rm s}(l_{\rm s})$ in \eqref{eq:Ys_backshort}.}

  \label{fig:sliding_short}
\end{figure}

A modified WR--42 waveguide fixture operating at
\(f_0=17.8~\text{GHz}\) was constructed to realize the integrated tuner in hardware (with two \emph{in-situ} tuning knobs $l_{\rm s}$ and $h$). Figure~\ref{fig:sliding_short} shows the
rectangular section with an adjustable sliding short that sets the stub offset \(l_{\rm s}\). The copper probe of diameter \(d=0.76~\text{mm}\) is mounted on a micrometer drive (resolution 25~\textmu m), enabling in situ adjustment of the exposed length \(h\). The inner conductor passes through a cylindrical doorknob post whose gap \(g\) and height \(l_{\rm d}\) were pre-machined to the nominal values predicted by Section~\ref{sec:probe_dknob}.

\begin{figure}[ht!]
  \centering
  \includegraphics[width=0.50\textwidth]{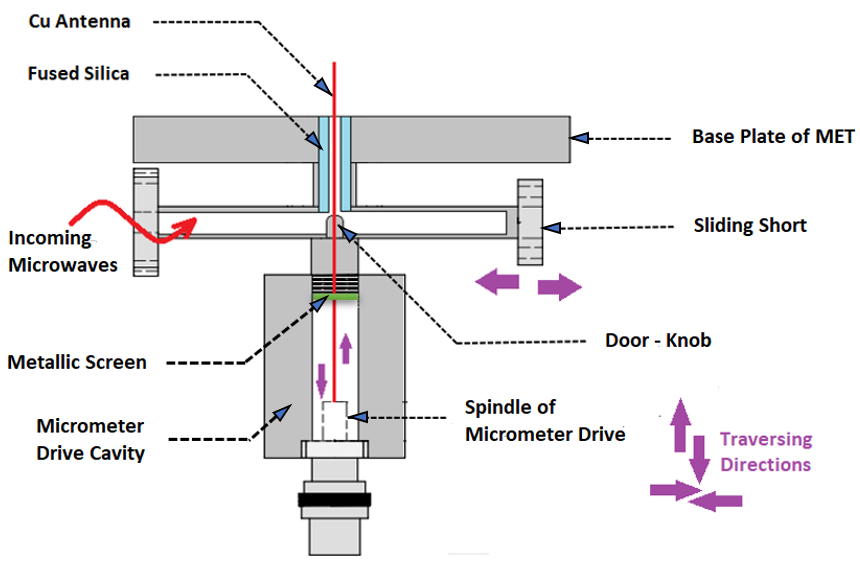}
  \caption{Cut-away of the antenna-sliding-short system.  The micrometer
    drive sets the exposed probe length \(h\); the plunger fixes
    \(l_{\rm s}\).  The doorknob post (right) realizes \(L_{\rm d}^{\star}\) and
    \(C_{\rm d}\).}
  \label{fig:antenna_assembly}
\end{figure}

Figure~\ref{fig:antenna_assembly} details the probe assembly. A
1~mm-thick fused-silica sleeve insulates the center conductor from the housing, providing the shunt capacitance \(C_{\rm fs}=0.06~\text{pF}\) included in the analytical model. The feedthrough capacitance was estimated using a quasi-static coaxial approximation,
$C_{\mathrm{fs}}\approx \dfrac{2\pi\epsilon_{0}\epsilon_{\rm r}\,\ell_{\mathrm{fs}}}{\ln(b/a)}$, with $\epsilon_{r}\approx 3.8$ for fused silica, $a=d/2$ the probe radius, $b$ the inner radius of
the surrounding housing, and $\ell_{\mathrm{fs}}$ the axial length of the fused-silica section, yielding $C_{\mathrm{fs}}\approx 0.06$ pF at Ku-band. The doorknob step is dimensioned so that \(L_{\rm d}^{\star}(l_{\rm d},h_{\rm g})\) in \eqref{eq:Ld_star} satisfies the real-part constraint in \eqref{eq:match_conditions}. The WR--42 fixture (Figs.~\ref{fig:sliding_short}--\ref{fig:antenna_assembly})
serves as both (i) the launch adapter for the MET cavity and (ii) a
general-purpose integrated waveguide-to-coax tuner for characterizing probe-coupled high-$Q$ resonators. It also provides a convenient platform for verifying tuner models with VNA measurements.

\begin{figure}[ht!]
  \centering
  \includegraphics[width=0.30\textwidth]{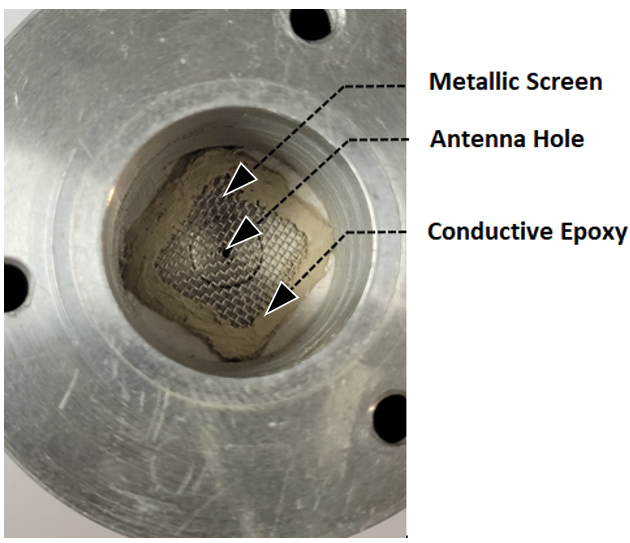}
  \caption{Metallic screen closing the micrometer cavity.  Experimental evidence shows \(>10\;\text{dB}\) suppression of spurious leakage at 17.8 GHz.}
  \label{fig:metal_screen}
\end{figure}

Without shielding, the coaxial feed path beneath Plane~B can excite
parasitic fields inside the micrometer-drive cavity and distort the measured reflection coefficient $|S_{11}|$. A 150-\textmu m stainless-steel mesh epoxied across the threaded bore (Fig.~\ref{fig:metal_screen}) serves as an RF seal that suppresses this leakage while leaving the mechanical drive unaffected. With the mesh installed, spurious leakage features near 17.8~GHz are reduced by more than 10~dB; the added capacitive loading is estimated to be $<0.03$~pF and is negligible compared to \(C_{\rm fs}\).

\begin{figure}[ht!]
  \centering
  \includegraphics[width=0.33\textwidth]{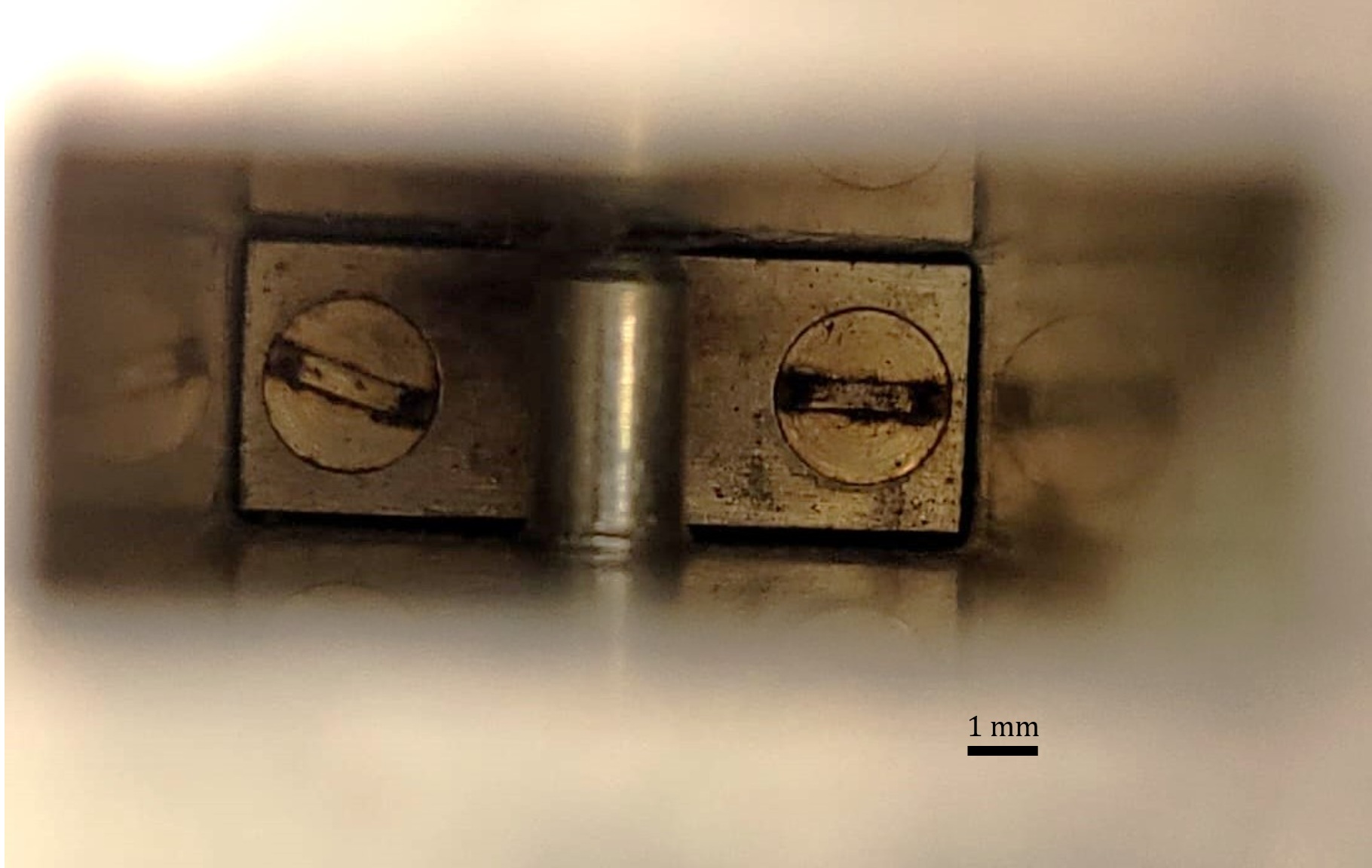}
  \caption{Doorknob post viewed along the guide axis.  Gap \(g\) and
    post height \(l_{\rm d}\) were set to the analytical optimum.}
  \label{fig:door_knob}
\end{figure}

\begin{figure}[ht!]
  \centering
  \includegraphics[width=0.5\textwidth]{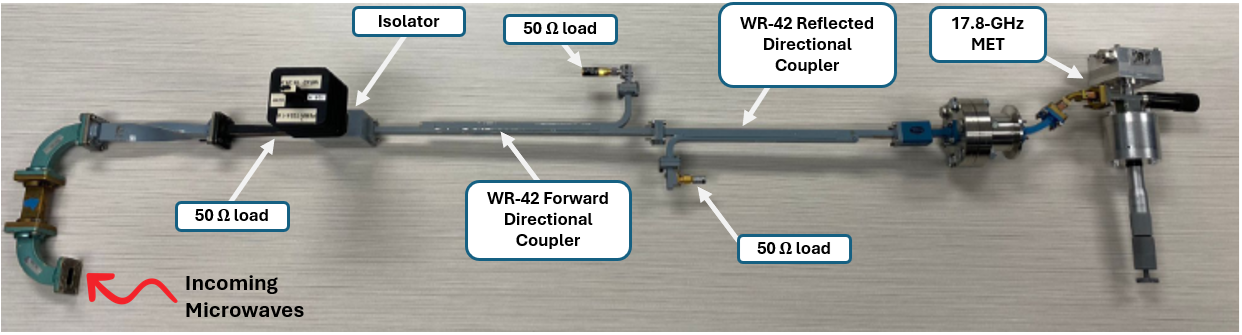}
  \caption{Setup of the transmission line assembly to the MET.}
  \label{fig:EntireSetup}
\end{figure}

A doorknob impedance transformer machined into the broad wall of the
waveguide (Fig.~\ref{fig:door_knob}) transforms the dominant-mode
(TE$_{10}$) wave impedance (440--640~$\Omega$) to the 50~$\Omega$ coaxial probe feed, maximizing power transfer in accordance with the synthesis in Section~\ref{sec:probe_dknob}.
S-parameters were recorded with a Keysight N5224B PNA network analyzer after TRL calibration referenced to the WR--42 flange (Plane~A). TRL and multiline-TRL are standard self-calibration approaches for accurate reference-plane definition in broadband fixtures \cite{EngenHoer1979TRL,Marks1991MultilineTRL,Kim2001,Papapolymerou2003}. An isolator and WR--42 directional couplers protected the analyzer during plasma ignition, as shown in Fig.~\ref{fig:EntireSetup}. The loaded quality factor $Q_{\rm L}$ was extracted from the measured reflection data using the absorbed-power spectrum $P_{\mathrm{abs}}/P_{\mathrm{inc}} = 1 - |S_{11}|^{2}$. We computed the full-width at half-maximum bandwidth $\Delta f = f_{2} - f_{1}$ by finding the frequencies where
$P_{\mathrm{abs}}(f_{1,2}) = \frac{1}{2} P_{\mathrm{abs,max}}$, and then determined $Q_{L} = f_{0}/\Delta f$. The uncertainty in \(|S_{11}|\) is \(\pm 0.15~\text{dB}\) (95\% CI) after calibration residuals.

The backshort position $l_{\rm s}$ was swept over 0--15 mm to characterize the matching capabilities and identify parasitic mode limits (see Section~\ref{results}), though nominal operation is confined to $l_{\rm s} < 12$ mm. The probe insertion depth \(h\) was then adjusted in 50-\textmu m steps to align the critical-coupling point (\(\beta=1\)) predicted by \eqref{critical coupling}; convergence took fewer than four iterations due to the closed-form guidance. Finally, \(l_{\rm s}\) was fine-trimmed by \(<0.2\lambda_{\rm g}\) to achieve a loaded quality factor of \(Q_{\rm L}\approx 900\). The resulting mechanical resolution in \(h\) and \(l_{\rm s}\) is comparable to or finer than that used in recent tuner prototypes \cite{Kiuru2007,Shaffer2024}, enabling reproducible mapping between mechanical settings and electrical response.

\begin{figure}[h!]
  \centering
  \includegraphics[width=0.45\textwidth]{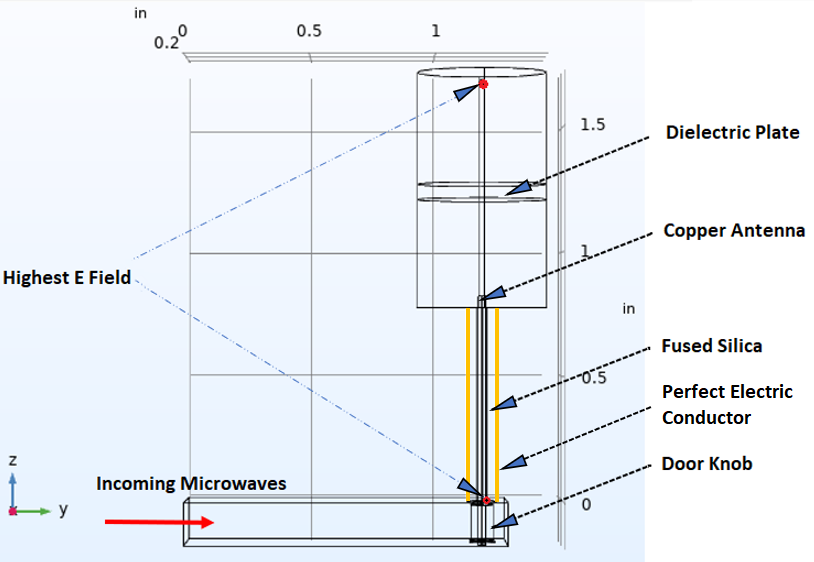}
  \caption{Computational domain of the MET with antenna, doorknob, and sliding-short arrangement.}
  \label{fig:compDomainMET}
\end{figure}

Full-wave simulations were performed in COMSOL Multiphysics to model and analyze the MET system with a sliding short, doorknob transition, and variable-height probe arrangement (Fig.~\ref{fig:tuner_circuit}). The computational domain is shown in Fig.~\ref{fig:compDomainMET}. The simulation objective was to minimize reflected power (i.e., minimize \(|S_{11}|\)) under no-plasma conditions while maximizing the electric-field magnitude near the nozzle for plasma ignition (Fig.~\ref{normEfieldMET}), with the standing-wave maximum centered near the probe location (Fig.~\ref{normEfieldWR42}). The Electromagnetic Waves, Frequency Domain interface was used with a single-mode TE$_{10}$ waveguide port excitation at the WR--42 flange, and S-parameters were computed directly from the port fields.

The computational domain was discretized using a tetrahedral mesh of
38{,}647 elements (average element quality 0.66). Local refinement was applied near the antenna (probe) tip, the fused-silica sleeve, and the doorknob step to resolve the highest field gradients. A mesh-convergence check with one level of uniform refinement produced $<2\%$ variation in the simulated reflection coefficient \(|S_{11}|\) at resonance, confirming mesh independence of the results.
\section{Results and Discussion} 
\label{results}
\begin{figure}[h!]
    \centering
    \includegraphics[width=0.4\textwidth]{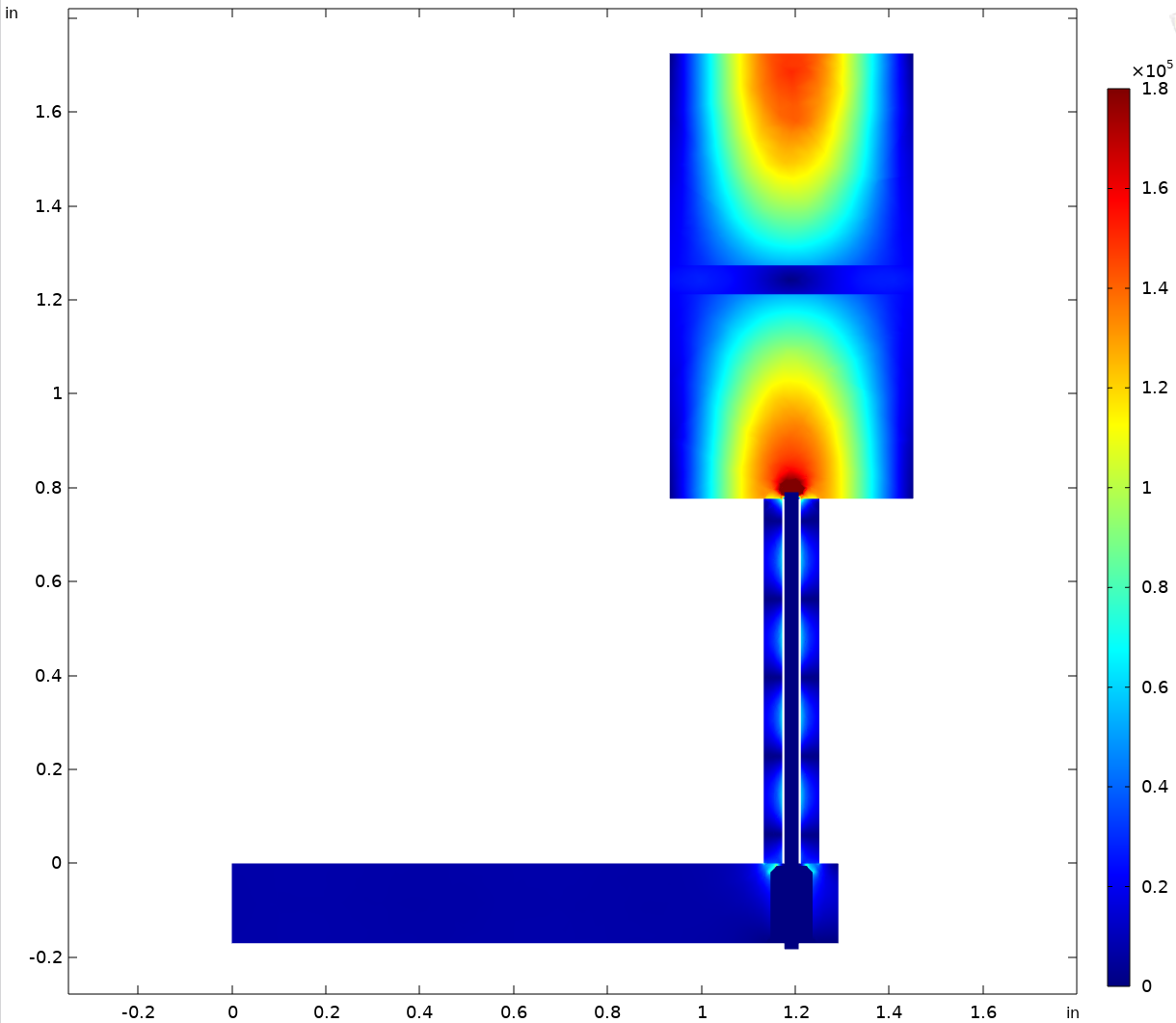}
    \caption{Simulated normalized electric-field magnitude $|E|$ in the MET chamber for the tuned configuration ($f=18.14$~GHz, $h=0.55$~mm, $l_{\mathrm{s}}=0.80$~mm). Fields are normalized to 1~W incident power.}
    \label{normEfieldMET}
\end{figure}

Electromagnetic simulations of the MET chamber (computational domain in Fig.~\ref{fig:compDomainMET}) identify the configuration in Fig.~\ref{normEfieldMET} as yielding the largest electric-field magnitude in the nozzle region. 

\begin{figure}[h!]
    \centering
    \includegraphics[width=0.45\textwidth]{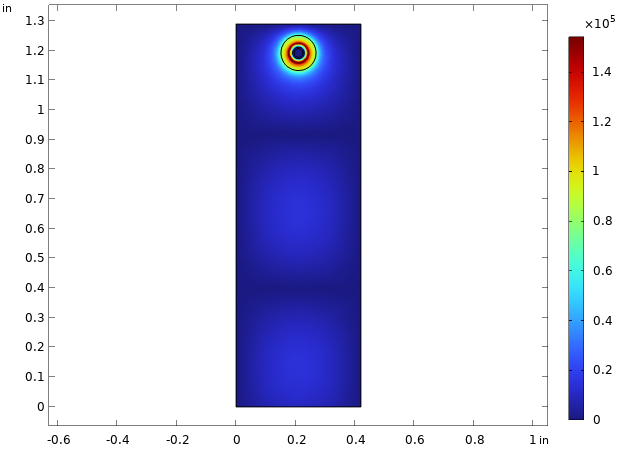}
    \caption{Top view of the WR-42 feed section showing the normalized electric-field magnitude $|E|$ for the tuned configuration in Fig.~\ref{normEfieldMET}.}
    \label{normEfieldWR42}
\end{figure}

For the tuned setting---antenna (probe) height $h=0.55$~mm, backshort distance $l_{\mathrm{s}}=0.80$~mm, and frequency $f=18.14$~GHz---driven with $P_{\mathrm{in}}=1$~W, the simulated peak field reaches $|E| \approx 1.8\times10^{5}$~V/m both near the nozzle and at the WR-42 feed where the antenna is located (Figs.~\ref{normEfieldMET} and \ref{normEfieldWR42}). Unless otherwise stated, fields shown are normalized to 1~W incident power.

\begin{figure}[h!]
    \centering
    \includegraphics[width=0.45\textwidth]{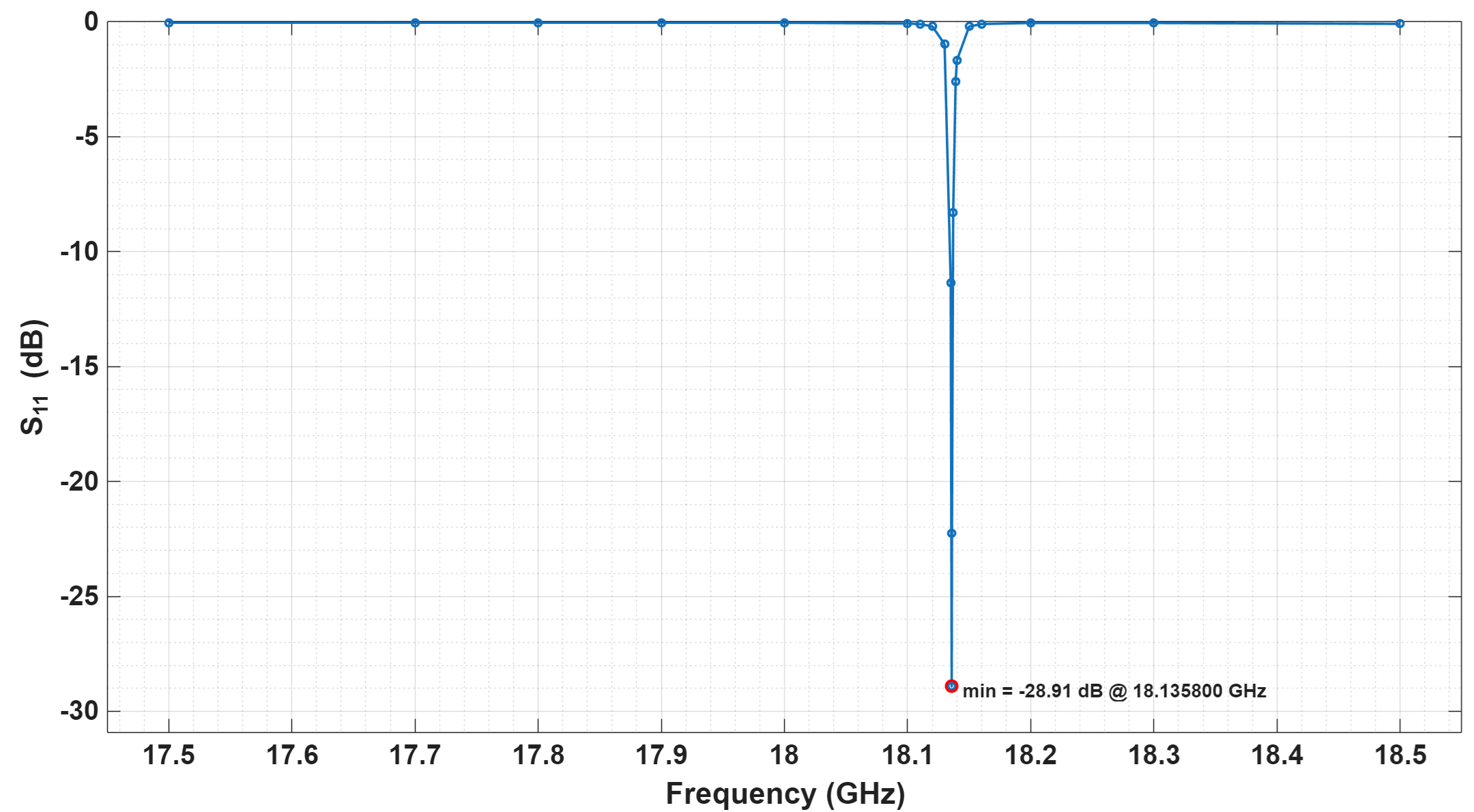}
    \caption{Simulated reflection coefficient $S_{11}$ (dB) versus frequency for the tuned configuration
    ($h = 0.55$~mm, $l_{\rm s} = 0.80$~mm). A minimum near $f = 18.14$~GHz reaches $\sim -30$~dB.}
    \label{S11-COMSOL}
\end{figure}

\begin{figure}[h!]
        \centering
        \includegraphics[width=0.4\textwidth]{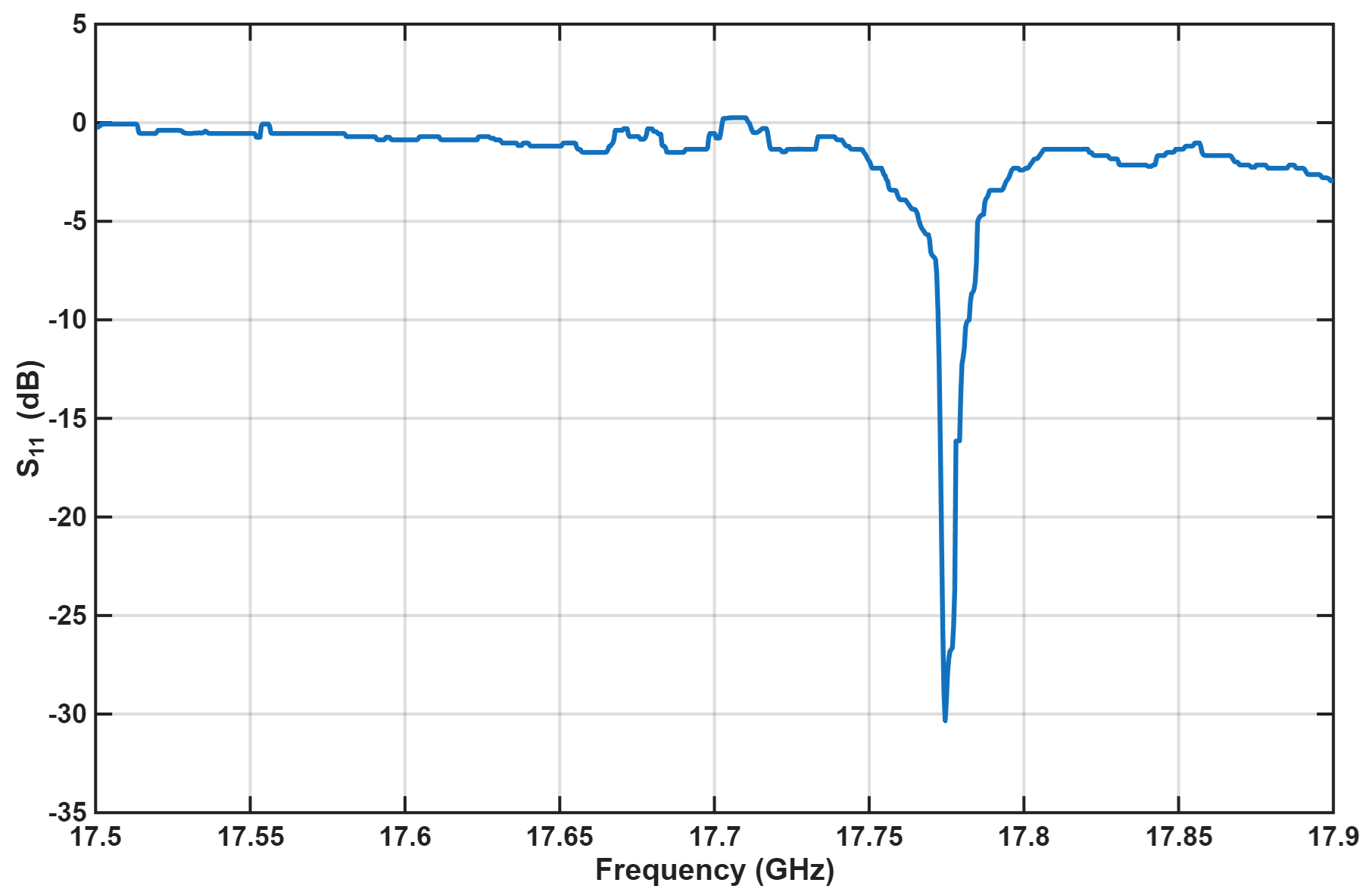}
        \caption{$S_{11}$ value of $\sim -30$~dB at a resonant frequency of 17.775 GHz.}
        \label{S11-Exp}
\end{figure}

The tuned case exhibits a simulated input reflection magnitude of $|S_{11}|\approx -30$~dB (i.e., $\mathrm{RL}\approx 30$~dB) at $f=18.14$~GHz (Fig.~\ref{S11-COMSOL}). The measured return-loss trace (Fig.~\ref{S11-Exp}) shows a comparable deep match with a minimum at $f_0=17.775$~GHz and $|S_{11}|\approx -30$~dB ($\mathrm{RL}\approx 30$~dB), confirming that the interface can be tuned to near-critical coupling in hardware. The $\approx 2\%$ frequency offset between the representative FEM case and measurement is attributed to a combination of machining tolerances, assembly stack-up, and effective dielectric/loading differences that shift
the high-$Q$ cavity resonance; small residual error from the first-order probe-reactance approximation in \eqref{Xp} may also contribute.

A coarse parameter study over antenna heights $h\in[0.4,\,1.2]$~mm and frequencies $f\in[17.8,\,18.2]$~GHz, with the sliding-short position re-optimized at each point, reveals multiple settings that achieve $|S_{11}|$ between $-10$ and $-20$~dB while sustaining simulated peak nozzle fields on the order of $10^{5}$~V/m. For the remainder of this COMSOL Multiphysics discussion we adopt the representative tuned point ($h=0.55$~mm, $f=18.14$~GHz) due to its deeper match and stronger field localization.

\begin{figure}[h!]
    \centering
    \includegraphics[width=0.45\textwidth]{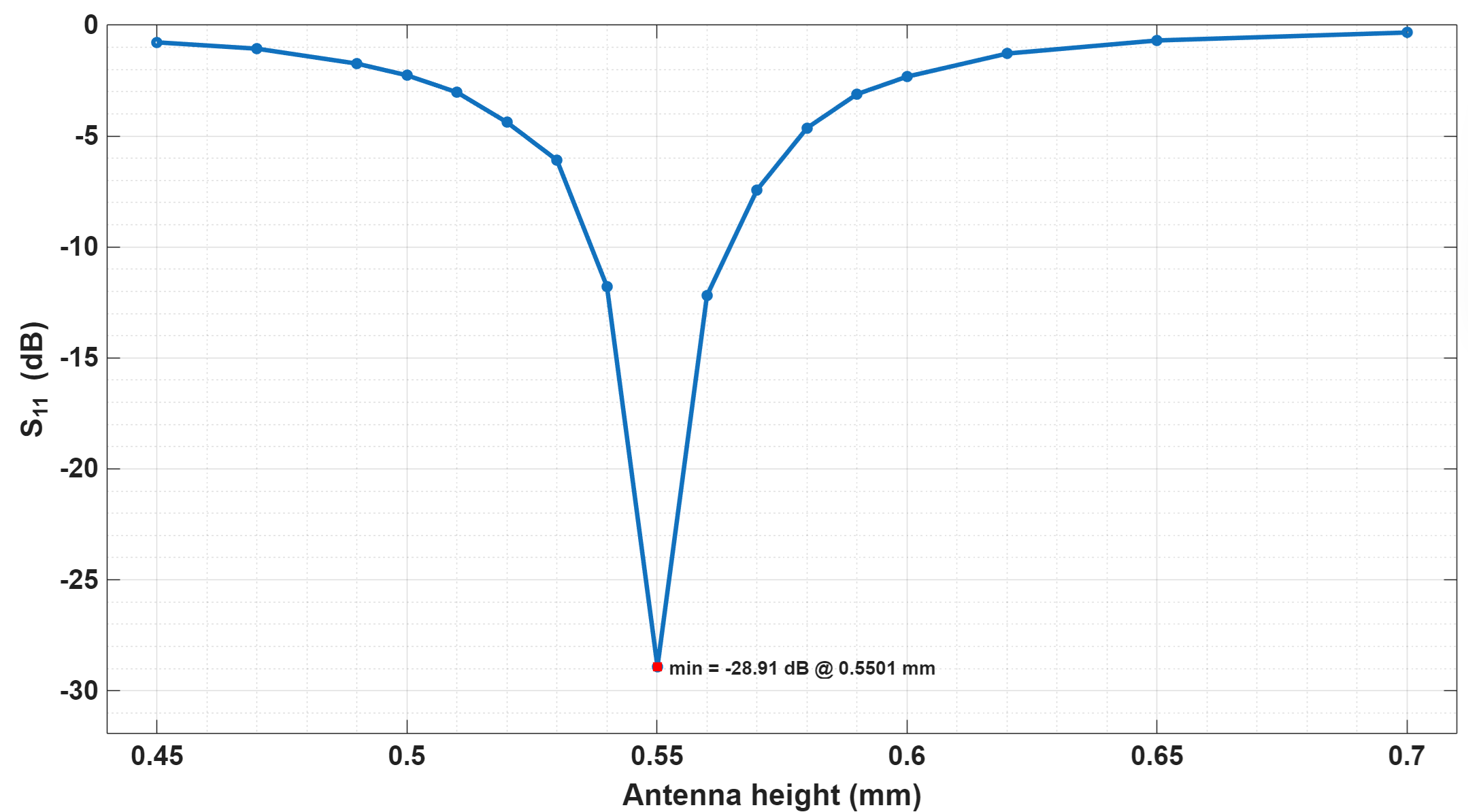}
    \caption{Sensitivity of $S_{11}$ (dB) to antenna height: simulated $S_{11}$ versus $h$ at fixed
    $f = 18.14$~GHz and $l_{\rm s} = 0.80$~mm. An optimum occurs near $h = 0.55$~mm.}
    \label{AntHeight}
\end{figure}

To quantify sensitivity, we fixed $f=18.14$~GHz and $l_{\mathrm{s}}=0.80$~mm and swept $h$. The results (Fig.~\ref{AntHeight}) show a pronounced optimum near $h=0.55$~mm, confirming that probe height is a critical control parameter for simultaneously minimizing $S_{11}$ and maximizing the local field in the nozzle region.

\begin{figure}[h!]
    \centering
    \includegraphics[width=0.45\textwidth]{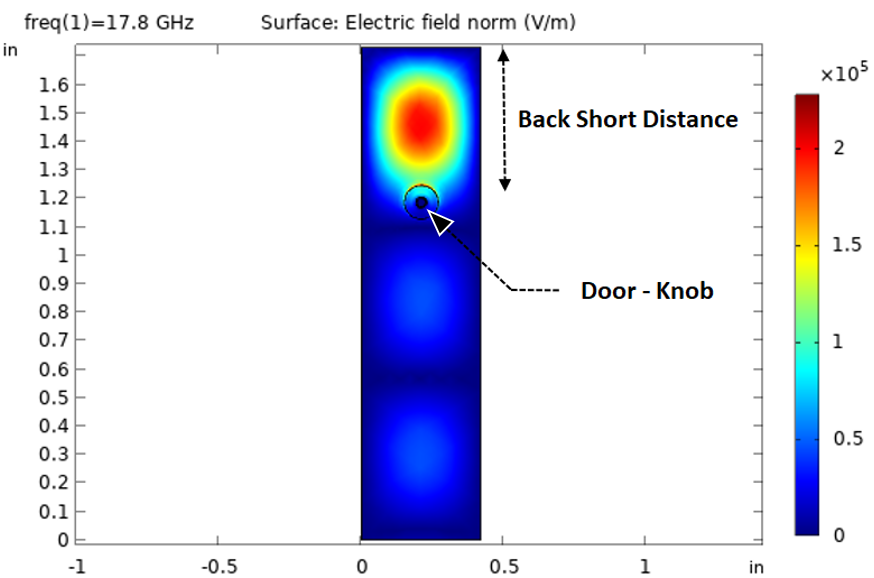} 
    \caption{Top view of the WR-42 feed showing a strong standing wave behind the doorknob when $l_{\rm{s}}>12$~mm (parasitic stub resonance). Fields are normalized to 1-W incident power. Frequency of 17.8 GHz was simulated since the traveling wave tube amplifier used in the experiment could only go up to 18 GHz. }
    \label{fig:backshort_field}
\end{figure}

Increasing the waveguide backshort length beyond $l_{\rm{s}}\!\approx\!12$~mm creates a resonant section behind the doorknob that supports a strong standing wave as shown in Fig.~\ref{fig:backshort_field}. This can reduce port reflection (i.e., improve $|S_{11}|$) while localizing a significant fraction of the electromagnetic energy in the backshort region rather than in the MET cavity. Because $S_{11}$ only quantifies the power returned to the source, it does not reveal where the nonreflected power is stored or dissipated; field maps are needed to determine whether the energy is actually delivered to the nozzle region.

\begin{figure}[h!]
    \centering
    \includegraphics[width=0.45\textwidth]{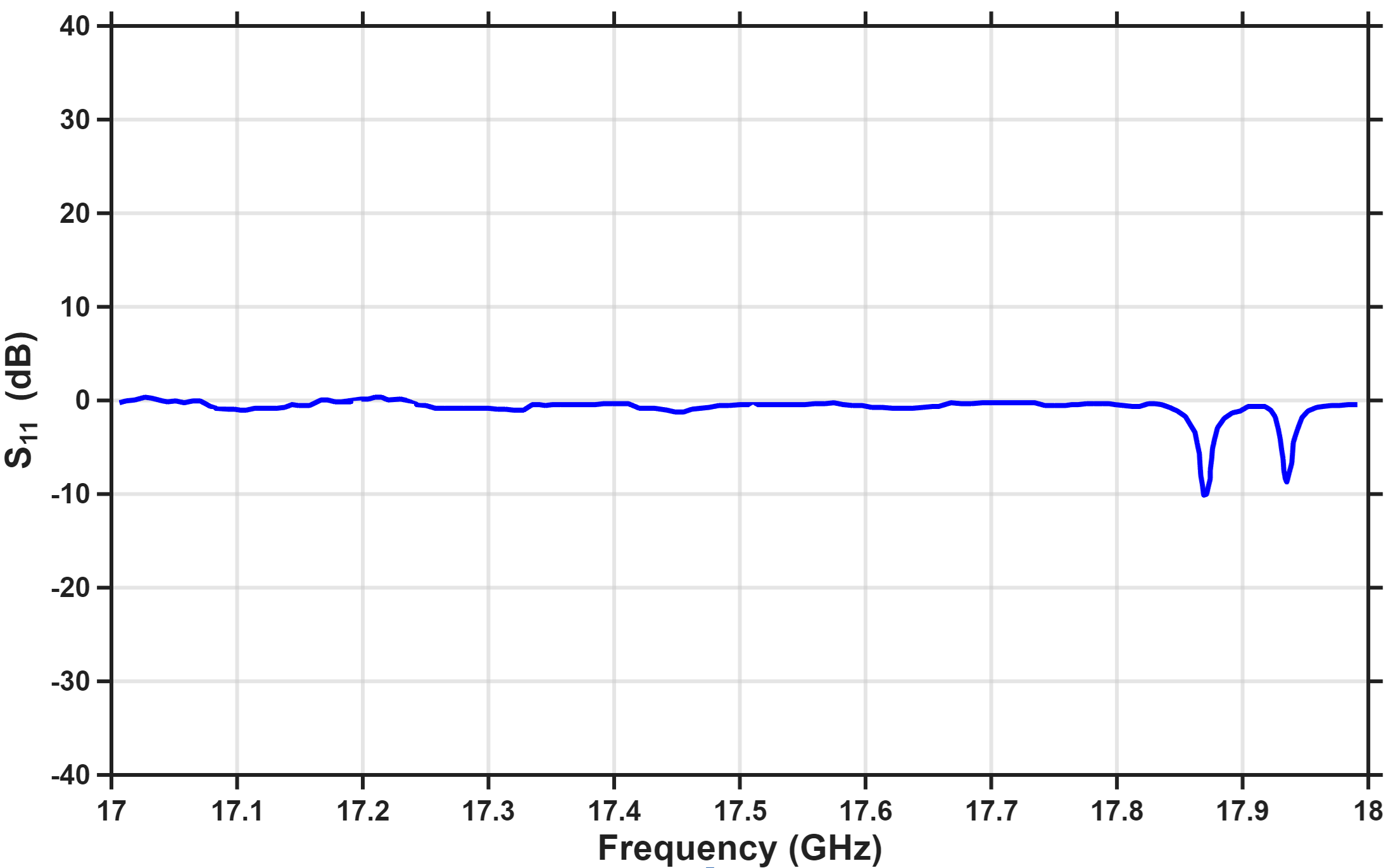}
    \caption{Measured reflection coefficient $S_{11}$ (dB) for $l_{\rm s} > 12$~mm showing two minima:
    $-8.95$~dB at 17.935~GHz and $-10.58$~dB at 17.87~GHz. The lower-frequency notch is associated
    with the MET cavity; the higher-frequency notch stems from the backshort stub.}
    \label{fig:s11_twin_trough}
\end{figure}

Figure \ref{fig:s11_twin_trough} shows the measured VNA response for the simulated field build-up behind the doorknob for $l_{\rm{s}}>12$~mm. Two distinct minima appear in $S_{11}$ at $f\!=\!17.935$~GHz and $f\!=\!17.87$~GHz with $S_{11}\!=\!-8.95$~dB and $-10.58$~dB, indicating a two-resonator system: a backshort stub resonance and the MET cavity resonance.

The leftward drift of the parasitic notch with increasing $l_{\rm{s}}$ follows the guided-wavelength condition for the dominant waveguide mode,
\begin{equation}
    \lambda_{\rm g}(f)=\frac{\lambda_0}{\sqrt{1-(f_{\rm c}/f)^2}}, \quad f_{\rm c}=\frac{c}{2a}\,,
\end{equation}
so that the stub resonance near $m\lambda_{\rm g}/2\approx l_{\mathrm{s}}$ occurs at lower $f$ as $l_{\rm{s}}$ grows. For WR-42 ($a=10.668$~mm, $f_{\rm c}\!\approx\!14.06$~GHz), $\lambda_{\rm g}\!\approx\!26.7$~mm at 18~GHz; thus $l_{\rm{s}}\!=\!12$--14~mm corresponds to $\sim$0.45--0.52$\lambda_{\rm g}$, which explains the strong field localized behind the doorknob (Fig.~\ref{fig:backshort_field}) and the twin-trough response (Fig.~\ref{fig:s11_twin_trough}). When the parasitic stub mode strengthens, it can detune and effectively suppress coupling to the MET cavity, degrading field intensity where it is needed.

From a design standpoint, low-power METs and surface-wave sources should either keep $l_{\rm{s}} \leq 0.4\lambda_g$ at the operating frequency or incorporate damping/irising to suppress the backshort stub, ensuring that a deep $S_{11}$ minimum coincides with strong field localization in the nozzle region rather than in the backshort. In the language of resonant impedance tuners, the backshort region effectively forms an unintended second resonator in cascade with the cavity \cite{Shaffer2024}, which explains the twin‑notch $|S_{11}|$ response.

With the measurement chain of Fig.~\ref{fig:EntireSetup}, the measured reflection coefficient in Fig.~\ref{S11-Exp} shows
a minimum at $f_0 = 17.775$~GHz with $|S_{11}| \approx -30$~dB
(i.e., $\mathrm{RL} \approx 30$~dB), confirming efficient matching at the operating frequency.

Table~\ref{tab:beta_ql} summarizes the coupling coefficient $\beta$ and loaded quality factor $Q_{\rm L}$ derived from the analytical model, COMSOL simulations, and experimental measurements, demonstrating excellent agreement with the critical-coupling target. Because \eqref{eq:QL_S11} implies $|S_{11}(f_{0})|=|(\beta-1)/(\beta+1)|$, the magnitude alone admits reciprocal solutions $\beta$ and $1/\beta$ (over- or under-coupled). While the phase trajectory on the Smith chart can theoretically resolve this ambiguity (by determining if the resonance locus encloses the origin), the measured deep match  ($|S_{11}(f_{0})|\approx -30~\mathrm{dB}$) implies $|\beta-1|\lesssim 0.07$. Consequently, the distinction between the two solutions ($\beta \approx 1.065$ or $0.939$) is negligible, and the system is effectively near-critical for practical power transfer.
\begin{table}[h!]
\caption{Loaded quality factor and coupling coefficient at the tuned operating point. 
$\beta$ is inferred from $|S_{11}(f_0)|=|\frac{\beta-1}{\beta+1}|$.}
\label{tab:beta_ql}
\centering
\footnotesize
\setlength{\tabcolsep}{4pt}
\renewcommand{\arraystretch}{1.15}
\begin{tabular}{lcc}
\hline \hline  
Source & $\beta$ & $Q_{\rm L}$ \\
\hline
ABCD model (critical-coupling target) & 1.00 & $\approx 900$ \\
FEM tuned point ($|S_{11}|\approx-30$ dB) & 0.94 or 1.07 & --- \\
Measured tuned point ($|S_{11}|\approx-30$ dB) & 0.94 or 1.07 & $\approx 900$ \\
\hline \hline 
\end{tabular}
\end{table}

\begin{figure}[h!]
        \centering
        \includegraphics[width=0.4\textwidth]{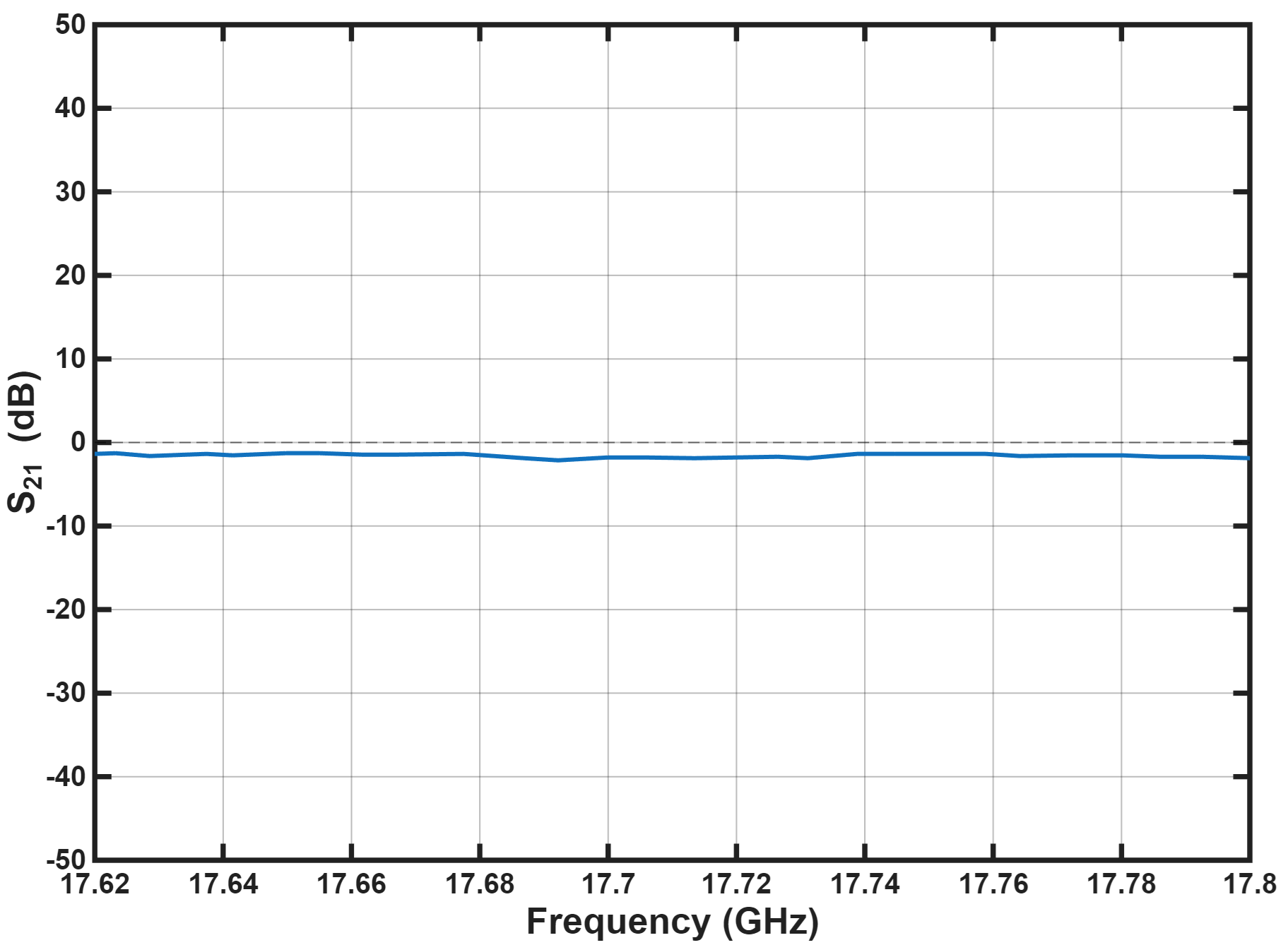}
        \caption{Measured transmission coefficient ($S_{21}$) of the WR-42 feed network assembly (isolator, couplers, and waveguides) showing an insertion loss of $\approx 0.8$~dB near 17.75~GHz.}
        \label{S21-Exp}
\end{figure}

Feed-network loss was quantified by measuring $S_{21}$ of the waveguide assembly (isolator, couplers, sections) separately. Because the MET cavity with the integrated tuner operates as a one-port termination, its primary characterization relies on $S_{11}$ referenced to Plane~A (Fig.~\ref{fig:tuner_circuit}). To isolate the losses in the feed network, the WR-42 transmission-line assembly shown in Fig.~\ref{fig:EntireSetup} (comprising the isolator, directional couplers, and connecting waveguide sections) was characterized separately as a two-port device with the MET removed. In this configuration, Port~1 is defined at the ``incoming microwaves'' flange (Fig.~\ref{fig:EntireSetup}) and Port~2 at the flange that mates to the MET input plane (Plane~A). The reported $S_{21}$ therefore represents the insertion loss of the upstream transmission line assembly only. The measured transmission corresponds to an insertion loss of approximately 0.8~dB ($|S_{21}| \approx -0.8$~dB, Fig.~\ref{S21-Exp}) at 17.75~GHz.

\begin{table*}[t]
\centering
\caption{Comparison with Recent Impedance Tuners and Reconfigurable Matching Approaches}
\label{tab:prior_art}

\renewcommand{\arraystretch}{1.2}
\setlength{\tabcolsep}{3pt}

\begin{tabularx}{\textwidth}{l L c c c L L}
\toprule
\textbf{Ref.} & \textbf{Tuning Approach} & \textbf{Band} & \textbf{Min. Loss} & \textbf{Power} & \textbf{Coverage / Objective} & \textbf{Notes}\\
\midrule

This work
& Sliding short + adj. probe integrated at waveguide--cavity interface
& 17.7--18.1~GHz
& $\eta_{\rm coupling}$\textsuperscript{a}
& 10--81~W\textsuperscript{1}
& Critical coupling to high-$Q$ cavity load ($\beta \approx 1$)
& In-situ mechanical retuning (match \& coupling) \\

\addlinespace[4pt]

Shaffer \textit{et al.}~\cite{Shaffer2024}
& Coupled-resonator resonant tuner
& 4--8~GHz
& 0.4~dB\textsuperscript{2}
& 100~W
& $\ge$90\% Smith-chart coverage
& 48~dB repeatability \\

\addlinespace[4pt]

Roessler \textit{et al.}~\cite{Roessler2023}
& Plasma-switch reconfigurable tuner
& 2--4~GHz
& 0.77~dB
& 20--70~W\textsuperscript{3}
& $\ge$30\% Smith-chart coverage
& 26.7~$\mu$s switching speed \\

\addlinespace[4pt]

Kabir \textit{et al.}~\cite{Kabir2025}
& Capacitive-tuned SIW evanescent-mode cavity (plasma jet)
& 2.66--2.94~GHz
& N/R
& few~W\textsuperscript{4}
& Resonant plasma-jet matching (tuned resonance)
& Frequency tuning; coupling not independently tuned \\

\bottomrule
\multicolumn{7}{p{\textwidth}}{\vspace{2pt}\scriptsize
\textsuperscript{a}One-port integrated launcher; standard two-port insertion loss is not defined. Efficiency is captured in coupling factor $\eta_{\mathrm{coupling}}$.
\par
\textsuperscript{1}Power corresponds to demonstrated plasma operation in this work:
He in-situ retuning at $P_{\rm in}=10$~W, NH\textsubscript{3} ignition/retuning cases at $P_{\rm fwd}=13$--20~W, and H\textsubscript{2} operation at $P_{\rm fwd}\approx 77$--81~W with $\eta_{\rm coupling}\approx 83\%$--$93\%$.
VNA $S$-parameters reported in this paper correspond to no-plasma bench characterization at atmospheric conditions (outside the vacuum chamber).
\par
\textsuperscript{2}Loss metric reported as minimum \emph{transducer loss} in \cite{Shaffer2024}.
\par
\textsuperscript{3}\cite{Roessler2023} reports 20~W for all frequencies/states and measured power handling up to 70~W (state-dependent).
\par
\textsuperscript{4}\cite{Kabir2025} reports operation with only a few watts input power for plasma-jet generation; N/R: not reported.
}
\end{tabularx}
\end{table*}

As summarized in Table~\ref{tab:prior_art}, unlike tuners optimized for broad Smith-chart coverage, this work targets in-situ critical coupling ($\beta\approx 1$) to a specific high-$Q$ cavity mode under a dynamically evolving plasma load, while keeping the matching network inside the launcher and eliminating external stub-tuner boxes.

\begin{figure}[h!]
    \centering
    \includegraphics[width=0.35\textwidth]{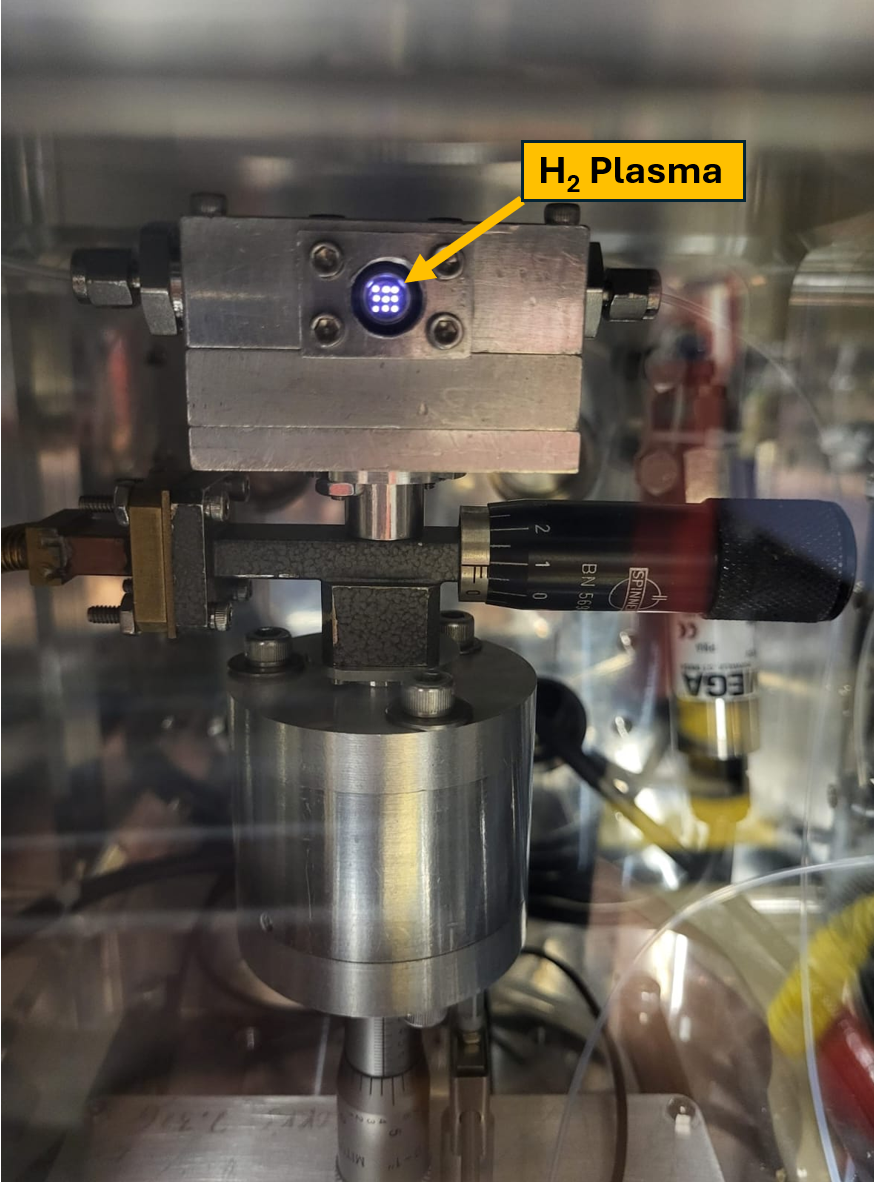}
    \caption{Representative plasma-on operation in the 17.8~GHz MET using the WR-42 feed with micrometer antenna and sliding short (propellant: H$_2$) inside a vacuum chamber. The same tuner geometry enables live re-tuning operation, $\eta_{\rm coupling}$, during discharge.}
    \label{H2-Plasma}
\end{figure}

We employed a WR-42 feed with a doorknob transition, micrometer-adjustable probe, and sliding short to retune the waveguide--cavity interface during plasma operation~\cite{BiswasIEPC2019,BiswasIEPC2022};
a representative H$_2$ discharge is shown in Fig.~\ref{H2-Plasma}. Electrically, the discharge behaves as a time-varying high-$Q$ load whose effective resistance and reactance drift with pressure and power~\cite{Ramesh2024,Ramesh2024_PlasmaResonator}.


\begin{figure}[h!]
    \centering
    \includegraphics[width=0.48\textwidth]{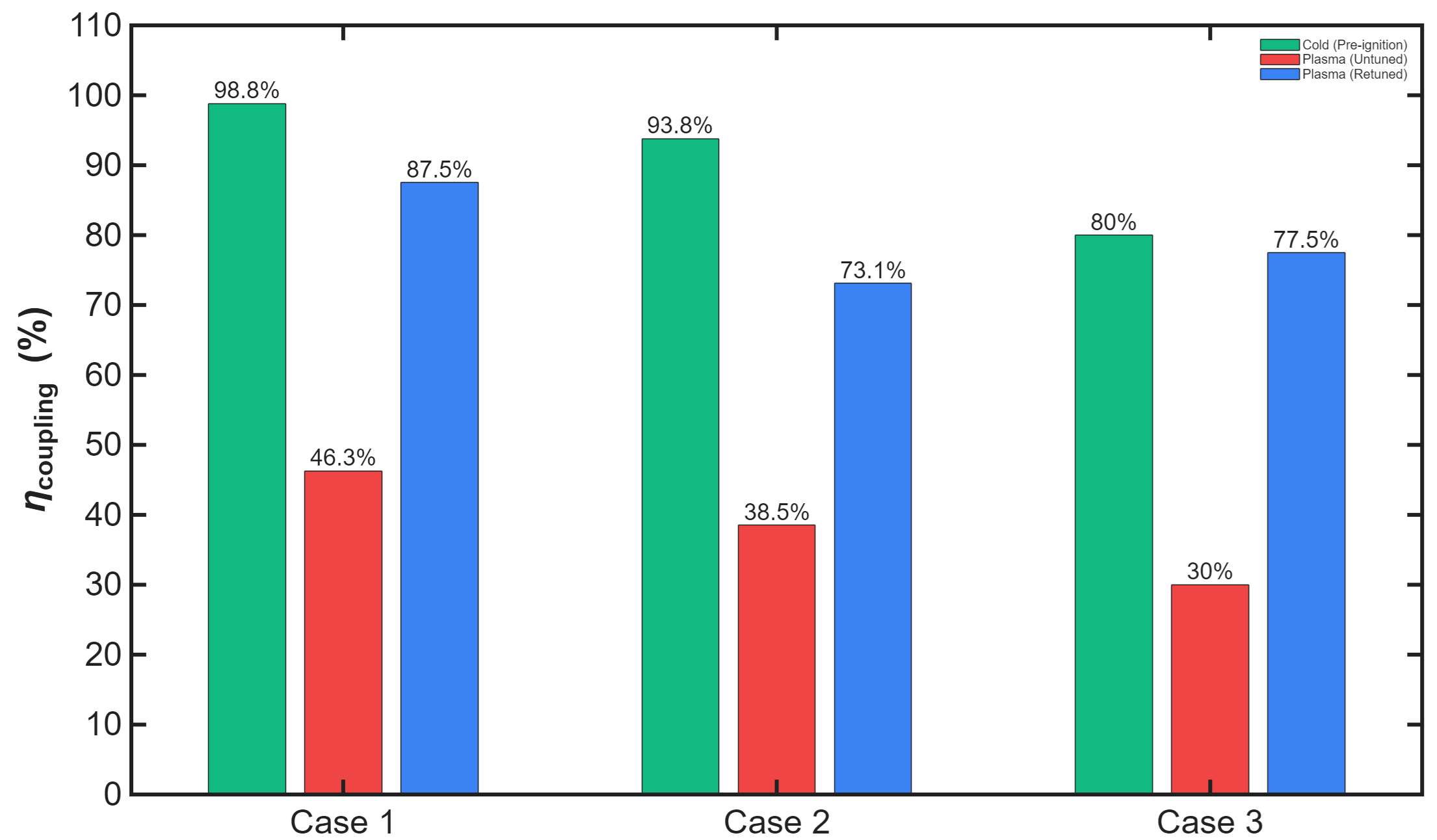}
    \caption{Comparison of coupling efficiency ($\eta_{\rm coupling}$) across three representative NH$_3$ operating points.
    In all cases, plasma ignition detunes the cavity (\textbf{Red Bar}) when held at the pre-ignition tuned setting.
    The cold (pre-ignition) \emph{tuned} setting is $(h_{\rm cold},\,l_{{\rm s},{\rm cold}})$ (\textbf{Green bar}) and the plasma-on \emph{re-tuned} setting is $(h_{\rm plasma},\,l_{{\rm s},{\rm plasma}})$ (\textbf{Blue Bar}), where $\Delta h=h_{\rm plasma}-h_{\rm cold}$ and
    $\Delta l_{\rm s}=l_{{\rm s},{\rm plasma}}-l_{{\rm s},{\rm cold}}$.
    \textbf{Case~1:} $f=17.70$~GHz, $(h_{\rm cold},l_{{\rm s},{\rm cold}})=(0.1~{\rm mm},\,7.575~{\rm mm})$, $(h_{\rm plasma},l_{{\rm s},{\rm plasma}})=(0.4$--$0.6~{\rm mm},\,7.875~{\rm mm})$ ($\Delta h\approx +0.3$--$0.5~{\rm mm}$, $\Delta l_{\rm s}=+0.3~{\rm mm}$).
    \textbf{Case~2:} $f=17.83$~GHz, $(0.5~{\rm mm},\,6.575~{\rm mm})\rightarrow(0.8$--$1.0~{\rm mm},\,5.275~{\rm mm})$ ($\Delta h\approx +0.3$--$0.5~{\rm mm}$, $\Delta l_{\rm s}=-1.3~{\rm mm}$).
    \textbf{Case~3:} $f=17.90$~GHz, probe-only retuning: $(0.6~{\rm mm},\,7.45~{\rm mm})\rightarrow(0.9$--$1.1~{\rm mm},\,7.45~{\rm mm})$ ($\Delta h\approx +0.3$--$0.5~{\rm mm}$, $\Delta l_{\rm s}=0$).}
    \label{Ammonia-Tuning}
\end{figure}

Plasma formation detunes the MET (Fig.~\ref{Ammonia-Tuning}); if the pre-ignition setting is held fixed, coupling collapses and the discharge self-quenches as mass flow $\dot m$ increases. Coupling is quantified at the tuner input as $\eta_{\rm coupling}=1-P_{\rm ref}/P_{\rm fwd}$ using directional couplers. In He tests at $P_{\rm in}=10$~W and $f\approx 17.78$~GHz, live adjustments of $(h,l_{\rm s})$ preserved a favorable match as the plasma impedance evolved, enabling $\dot m$ to be ramped from 25 to 350~sccm without extinction; over this sweep $\eta_{\rm coupling}$ increased from $\sim$43\% to $\sim$76\% and the stagnation-pressure ratio increased to $P_{0, {\rm h}}/P_{0, {\rm c}}\approx1.78$, consistent with stronger plasma heating.

Figure~\ref{Ammonia-Tuning} provides a compact NH$_3$ ignition demonstration at very low flow ($\dot{m}\approx 0.1$~sccm) and $P_{\rm fwd}=13$--20~W. In each case, the integrated tuner is
first adjusted before ignition (no plasma) to establish a high-coupling cold setting; holding this cold-tuned setting fixed after ignition produces a strong drop in $\eta_{\rm coupling}$, whereas
additional in-situ retuning during plasma-on operation restores coupling across the three operating points (e.g., from $\sim$80--99\% cold to $\sim$30--46\% plasma-untuned, then back to $\sim$73--88\% plasma-retuned). \noindent\textit{High-power H$_2$ operation:} Using the same directional-coupler power measurements at the tuner input, H$_2$ operation was demonstrated at $f\approx 17.8$~GHz (TM$^{z}_{011}$) with $h=0.61$~mm, $P_{\rm fwd}\approx 77-81$~W, and $\eta_{\rm coupling}\approx 83-93\%$ over $\dot{m}=100-196$~sccm.

These results contrast with cases tuned only pre-ignition, where coupling peaks and then collapses as $\dot{m}$ increases, ultimately extinguishing the discharge~\cite{Abaimov2015,BiswasIEPC2019}.
Accordingly, the integrated tuner (sliding short + probe with a fabrication-set doorknob transition) serves two roles: it establishes the initial cold-flow/no-plasma match to the cavity mode, and it provides plasma-on retuning via $(h,l_{\rm s})$ to track plasma-induced impedance drift, expanding the stable envelope to higher $\dot{m}$ and greater plasma heating at fixed input
power~\cite{BiswasIEPC2022}. This in-situ retuning is conceptually similar to adaptive/self-healing matching networks~\cite{Roessler2023,MoranGuizan2024,singh2021sixport}, but implemented mechanically at the waveguide--cavity interface rather than with switches or varactors. We view this primarily as an application example: the same integrated tuner (two \emph{in-situ} knobs) could provide real-time matching for other time-varying cavity loads, such as plasma jets~\cite{Kabir2025,Akram2025} or compact pulse-compressor cavities~\cite{Jiang2021}.

\section{Conclusion}
\label{conclusion}

We have demonstrated a compact sliding-short/probe tuner integrated into the launch adapter of a waveguide-coupled high-$Q$ cavity. The launcher combines a waveguide sliding short and a micrometer-adjustable coaxial probe as two \emph{in-situ} tuning knobs, with the doorknob transition geometry $l_{\rm d}$ fixed by fabrication. A closed-form chain-matrix model relates the geometric settings $(l_{\rm d},l_{\rm s},h)$ to the matching condition $\Gamma\!\rightarrow\!0$ and to the resonator loading parameters $\beta$ and $Q_{\rm L}$, while explicitly accounting for the fused-silica feedthrough capacitance. The resulting design rules were verified by full-wave simulations and S-parameter measurements on a WR-42 prototype, which achieved $|S_{11}|\approx -30$~dB near 17.7--18.1~GHz with $|S_{21}|\approx 0.7$--0.8~dB. Field maps show that the tuned configuration concentrates the electric-field maximum at the nozzle while avoiding spurious localization behind the doorknob.

A key design implication is that the backshort section is part of the resonant matching network. When $l_{\rm s}\gtrsim 0.5\lambda_{\rm g}$, a parasitic stub resonance forms behind the doorknob and produces a twin-notch $|S_{11}|$ response while storing energy in the backshort region rather than in the cavity load. Maintaining $l_{\rm s}\le 0.4\lambda_{\rm g}$ at the operating frequency avoids this unintended mode and ensures that a deep return-loss minimum coincides with field localization where it is required.

With plasma loading, the same in-situ knobs restore coupling as the discharge impedance evolves. Helium experiments at $P_{\rm in}=10$~W demonstrate stable operation while ramping the flow from 25 to 350~sccm with absorbed fraction increasing from $\sim$43\% to $\sim$76\%. NH$_3$ ignition tests at $\sim$0.1~sccm show that plasma onset detunes the cold-tuned match and that in-situ retuning recovers coupling across representative operating points. At $f\approx 17.8$~GHz, H$_2$ operation was demonstrated at $P_{\rm fwd}\approx 77-81$~W with $\eta_{\rm coupling}\approx 83-93$\% over $\dot m=100-196$~sccm. Together, these results show that a compact, mechanically actuated tuner integrated at the waveguide--cavity interface can provide high-power matching for time-varying high-$Q$ loads without external stub boxes, and can be scaled to other bands and applied to waveguide-coupled resonators in spectroscopy, plasma sources, and pulsed high-power microwave systems.

\section*{Acknowledgment}

This research was supported through a Graduate Student Teaching Assistantship from the Aerospace Engineering  Department at The Pennsylvania State University.




%





\ifCLASSOPTIONcaptionsoff
  \newpage
\fi





\bibliographystyle{IEEEtran}
\bibliography{IEEEabrv,Bibliography}

\vfill


\end{document}